\documentclass[preprint,amsmath,amssymb,aps,onecolumn]{revtex4-1}
\usepackage{graphicx}
\usepackage[colorlinks,linkcolor=black,anchorcolor=blue,citecolor=blue,urlcolor=blue]{hyperref}
\usepackage[right]{lineno}
\begin{document}
\title{Moir\'e-pattern evolution couples rotational and translational friction at crystalline interfaces}
\author{Xin Cao$^{1,\dagger}$}
\author{Andrea Silva$^{2,3,\dagger}$}
\author{Emanuele Panizon$^{4}$}
\author{Andrea Vanossi$^{2,3}$}
\author{Nicola Manini$^{5}$}
\author{Erio Tosatti$^{2,3,4}$}
\author{Clemens Bechinger$^{1}$}
\email{clemens.bechinger@uni-konstanz.de}
\thanks{$^{\dagger}$ Xin Cao and Andrea Silva contributed equally}

\affiliation{$^1$Fachbereich Physik, University Konstanz, 78464 Konstanz, Germany}
\affiliation{$^2$CNR-IOM, Consiglio Nazionale delle Ricerche - Istituto Officina dei Materiali, c/o SISSA,  34136, Trieste, Italy}
\affiliation{$^3$International School for Advanced Studies (SISSA), Via Bonomea 265, 34136 Trieste, Italy}
\affiliation{$^4$The Abdus Salam International Centre for Theoretical Physics (ICTP), Strada Costiera 11, 34151 Trieste, Italy}
\affiliation{$^5$Dipartimento di Fisica, Universit\`a degli Studi di Milano, Via Celoria 16, 20133 Milano, Italy\\}

\begin{abstract}
The sliding motion of objects is typically governed by their friction with the underlying surface. Compared to translational friction, however, rotational friction has received much less attention. Here, we experimentally and theoretically study the rotational depinning and orientational dynamics of two-dimensional colloidal crystalline clusters on periodically corrugated surfaces in the presence of magnetically exerted torques. We demonstrate that the traversing of locally commensurate areas of the moir\'e pattern through the edges of clusters, which is hindered by potential barriers during cluster rotation, controls its rotational depinning. The experimentally measured depinning thresholds as a function of cluster size strikingly collapse onto a universal theoretical curve which predicts the possibility of a superlow-static-torque state for large clusters. We further reveal a cluster-size-independent rotation-translation depinning transition when lattice-matched clusters are driven jointly by a torque and a force. Our work provides guidelines to the design of nanomechanical devices that involve rotational motions on atomic surfaces. 
\end{abstract}
\maketitle

\section{Introduction}
To set an object into motion typically requires a finite driving force to overcome the static friction with the surface underneath. Similarly, a finite torque must be applied to initiate a rotation. Although both effects originate from the same mechanisms, i.e. molecular adhesion and surface roughness \cite{bowden-tabor,persson}, the simultaneous translation and rotation of macroscopic objects demonstrate a nontrivial relation between static friction forces and torques \cite{dahmen2005pre}. Compared to macroscopic scales, where the overall tribological behaviour is usually explained in terms of time-honored, yet phenomenological, classical laws, the possible translation-rotation frictional interplay becomes physically much more intriguing, when dealing with atomically smooth crystalline contacts at the micro- and nanoscopic scales. These contacts appear in many nano-manipulation experiments and are crucial in micro- and nano-electro-mechanical systems (MEMS, NEMS) \cite{li2007nn,kim2007nt,bhushan2007me}. In such cases, friction strongly depends on the atomic commensurability of the surface lattices in contact \cite{marom2010prl}, which generate a rich tribological behavior including stick-slip motion and superlubric translational sliding \cite{falk2010nl,vanossi2013rmp,reichhardt2017rpp,vanossi2020nc,hod2018nat}. Contrary to translational nanofriction which received considerable experimental and theoretical attention during recent years, microscopic rotational friction has remained rather elusive despite being important for the reorientation dynamics and positioning of molecules and nano motors on atomic surfaces \cite{stipe1998sci,gimzewski1998sci,zheng2004jacs,delden2005nat,eelkema2009nat,manzano2009nm,filippov2009pre,tierney2011nn,pawlak2012an,schaffert2013nm,perera2013nn,simpson2019nc,jasper2020an}. In particular, it is unclear how rotational friction couples to the translational friction at atomic scales and how this depends on the properties of the two lattices in contact. This lack of knowledge is due to the difficulty of applying  well-controlled torques at nanoscopic length scales, but also results from the difficulty of the systematic variation of the lattice constant of materials. Such problems can be resolved by using micron-sized colloidal crystals sliding across patterned surfaces since torques and forces in such systems can be applied in a precise manner \cite{bohlein2012prl,libal2018nc,brazda2018prx,bililign2021np}. In addition, in such colloidal systems, contacts with almost arbitrary interface incommensurability can be created \cite{cao2019np,cao2021pre}.

Here, we experimentally and theoretically investigate the complex rotational motion of close-packed two-dimensional (2D) colloidal clusters which interact with a triangular surface lattice in presence of a constant external torque. We observe a non-monotonic contact-size dependence of the critical torque per particle required for rotational depinning of clusters when their lattice spacing differs from that of the substrate. We also discover a size-independent depinning boundary for clusters driven by a combination of external torques and forces. Our results are in excellent agreement with a theoretical model which considers the motion-induced evolution of the moir\'e pattern at the interface and coarse grains the locally commensurate moir\'e areas to Gaussian energy-density profiles. In contrast to its linear motion, the evolution of moir\'e pattern during rotation displays a qualitatively different and rather complex behavior: locally commensurate areas expand or shrink continuously in size, thus crossing the edges of the clusters, which is crucial for the depinning. Interestingly, our theoretical evaluation of the rotational depinning threshold reveals a super low-static-torque state which may find use for the engineering of low-friction nanomechanical gears. 

\begin{figure*}[t]
  \centering
  \includegraphics[width=1.0\columnwidth]{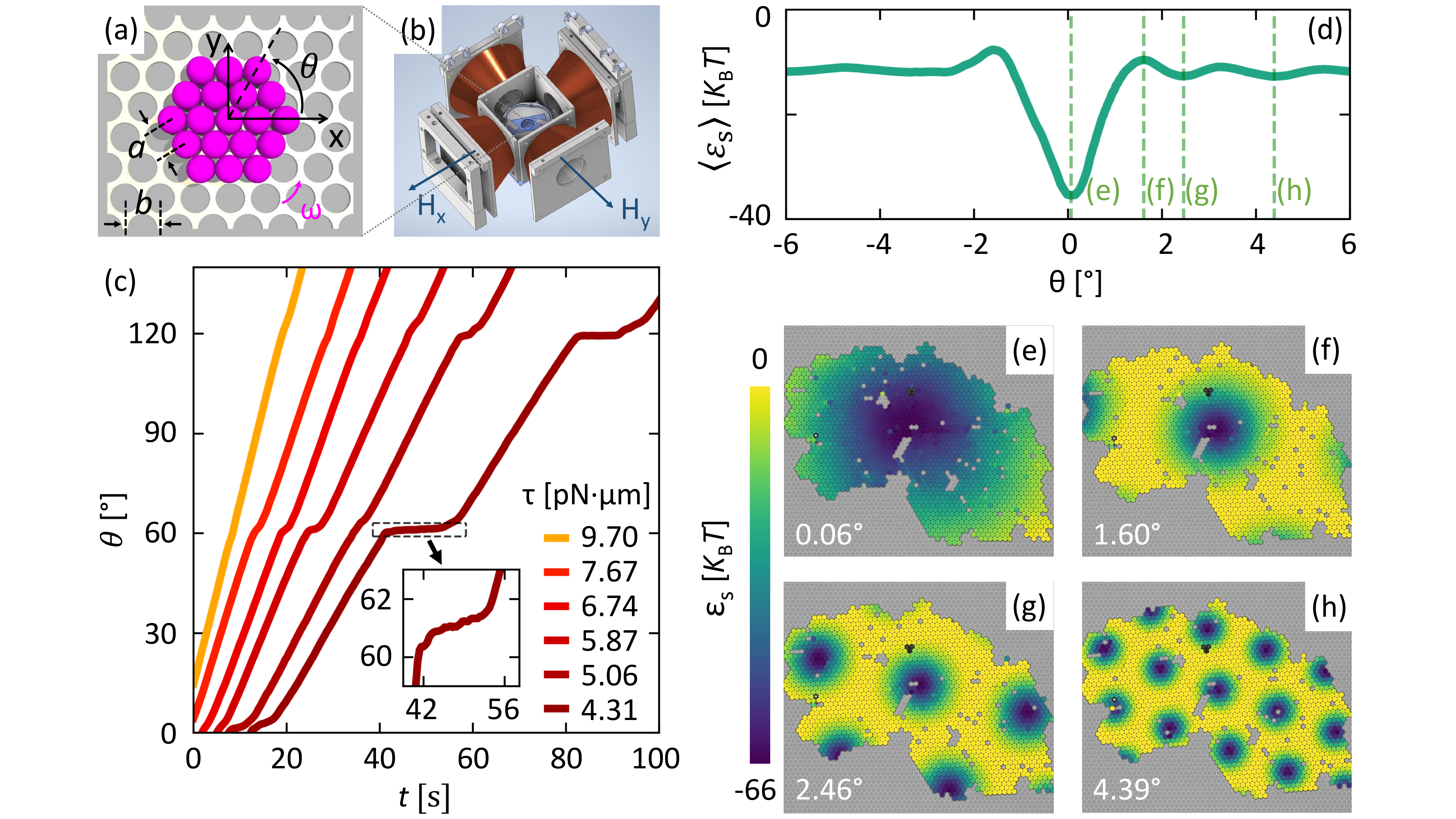}
\caption{Orientational cluster motion on periodic surfaces. (a,b) Illustration of the experimental setup. Two perpendicular pairs of coils generate a rotating magnetic field in the sample plane with frequency $\omega_\mathrm{H}$. It applies a torque $\tau$ to the clusters leading to their rotation with angular velocity $\omega\ll\omega_\mathrm{H}$ on top of a topographically patterned surface with triangular symmetry. The conical shape of the coils maximizes the magnetic strength and the area of the uniform field at the center. The angle $\theta$ denotes the lattice direction of the colloidal cluster, relative to the $x$ axis (i.e. substrate lattice direction). (c) Temporal evolution $\theta(t)$ for a cluster with $N = 133$ when rotating on a surface with $\delta=1.1\%$ under various applied torque $\tau$. Inset: Magnified view of the plateau on the $ \tau= 4.31~\mathrm{pN}\cdot\mu\mathrm{m}$ curve. (d) The interaction energy $\langle\epsilon_\mathrm{s}\rangle$ as a function of $\theta$ for a cluster with $N = 1715$ on a $\delta=1.1\%$ surface. $\langle~\rangle$ denotes the average over all particles in the cluster. The dashed lines indicate the snapshots in (e-h). (e-h) Snapshots of the cluster in (d) at angles $\theta=0.06^{\circ},~ 1.60^{\circ},~ 2.46^{\circ},~ 4.39^{\circ}$ under applied torque $\tau = 6.68~\mathrm{pN}\cdot\mu\mathrm{m}$. Particles are color coded according to their interaction energy $\epsilon_\mathrm{s}$ with the substrate (see Appendix \ref{appendmicromodel} for calculation of $\epsilon_\mathrm{s}$).}
\label{fig1}
\end{figure*}

\section{Experiments and results}
\subsection{Experimental sample preparation and torque realization}
Colloidal clusters are made from an aqueous suspension of superparamagnetic colloidal particles (diameter $\sigma= 4.45~\mu\mathrm{m}$) where a small amount (0.02\% in weight) of polyacrylamide (PAAm) is added. The PAAm causes strong interparticle bonds leading to rigid 2D clusters with the lattice constant $a=\sigma$ which is fixed in our experiments. Owing to their fabrication process, the clusters have a broad distribution in size and shape. As illustrated in Fig. \ref{fig1}(a), the clusters are interacting with a periodically corrugated substrate fabricated by photolithography. In our experiments we use four different substrate lattice spacings $b = 4.4,~4.6,~4.7,~4.8~\mu\mathrm{m}$, producing lattice-spacing mismatches $\delta=|1-a/b|=1.1\%,~3.3\%,~5.3\%,~7.3\%$, respectively. Application of a torque to the clusters is achieved by two mutually perpendicular pairs of coils [Fig. \ref{fig1}(b)] which create a magnetic field with components $H_x = H \cos\omega_\mathrm{H}t$ and $H_y = H \sin\omega_\mathrm{H}t$. This leads to rotation of the total $\mathbf{H}$ vector in the $x-y$ sample plane. The frequency $f_\mathrm{H} = \omega_\mathrm{H}/ (2\pi)$ was set to $f_\mathrm{H}= 10~\mathrm{Hz}$ in all measurements. The rotating $\mathbf{H}$ vector induces a rotating magnetization $\mathbf{M}_i$ within each superparamagnetic colloidal particle $i$ of a cluster. Due to a small phase lag in $\mathbf{M}_i$, the rotating magnetic field applies a torque $\Gamma = |\sum_i \mathbf{M}_i\times\mathbf{H}|$ to the entire cluster \cite{ranzoni2010lc,martinez2015pra}. This causes the cluster to rotate smoothly on top of a flat surface with an angular velocity $\omega=\tau/[\pi\eta a^3(1+I)]$ (see Video 1 and Fig. S1 of the Supplemental Material \cite{supplemental}). Here $\tau=\Gamma/N$ is the applied torque per particle, $N$ the number of particles, $N\pi\eta a^3(1+I)\omega$ the viscous torque of the cluster rotating in a liquid with viscosity $\eta$, and $I=\sum_i 3r_i^2/(N\sigma^2)$ a dimensionless factor which depends on the position of each particle $r_i$ relative to the rotation center. For details regarding sample preparation, cluster formation, particle tracking, and the calibration of the torque $\tau$ we refer to Appendix \ref{appendsample} and \ref{appendcalibration}. In the following we characterize the cluster’s angular velocity $\omega^* = (I+1)\omega = \tau/(\pi\eta a^3)$, which does not depend on the clusters’ size and shape.

\subsection{Orientational cluster motion}
The rotational dynamics of clusters is strongly modified in presence of a periodically patterned surface. This is illustrated in Fig. \ref{fig1}(c), which shows the time-dependence of the orientation $\theta$ of a cluster consisting of $N = 133$ particles rotating on a nearly matched surface ($\delta=1.1\%$) under various applied torques $\tau$. As expected, $\theta(t)$ displays an increasingly intermittent behavior for decreasing $\tau$, due to the increasing relative influence of the substrate corrugation. The interaction with the substrate leads to plateaus around high-symmetry angles $\theta = 0^{\circ}, 60^{\circ}, 120^{\circ}$ where the rotational velocity almost vanishes [inset of Fig. \ref{fig1}(c)]. The intermittent orientational dynamics originates from the rapidly changing cluster-substrate interaction energy $\langle\epsilon_\mathrm{s}\rangle$ near the high-symmetry angles. This is illustrated in Fig. \ref{fig1}(d). The reason for these energy oscillations are clarified in Fig. \ref{fig1}(e-h), reporting the local energy distribution for four snapshots of a cluster near $\theta = 0^{\circ}$. The low-energy spots (i.e. the dark-colored regions) arrange periodically on the cluster, forming the moir\'e pattern of the two contacting lattices. During rotation, and notably around $\theta = 0^{\circ}$, the low-energy moir\'e spots change drastically in size and spacing \cite{hermann2012jpcm}. As a consequence, they regularly move in and out of the cluster’s edge, as illustrated in Video 2 of the Supplemental Material \cite{supplemental}. The snapshot in Fig. \ref{fig1}(e) corresponds to a situation where the entire cluster is covered by a single, broad moir\'e spot which determines the absolute potential-energy minimum at $\theta = 0.06^{\circ}$ in Fig. \ref{fig1}(d). As the cluster rotates, the moir\'e spot shrinks and the potential energy increases. When the cluster rotates to $\theta = 1.60^{\circ}$, the potential energy reaches a maximum in Fig. \ref{fig1}(d) because neighbouring moir\'e spots have reached the edge of the cluster [Fig. \ref{fig1}(f)]. Upon further rotation, these neighbouring moir\'e spots move through the cluster’s edge, which leads to an energy local minimum in Fig. \ref{fig1}(d) at $\theta = 2.46^{\circ}$, where a first ring of moir\'e spots has moved inside the cluster [Fig. \ref{fig1}(g)]. Similarly, another energy local minimum arises at $\theta = 4.39^{\circ}$ when a second ring of surrounding moir\'e spots moves inside the cluster [Fig. \ref{fig1}(h)]. Video 3 provides an animation on how the moir\'e-pattern evolution determines the potential energy. These observations indicate that the oscillation of the potential energy near the high-symmetry angles depends strongly on the shape and size of the cluster. When the cluster size is similar or smaller compared to the size of the low-energy spots [Fig. \ref{fig1}(e)], the oscillation amplitude of the potential energy becomes large. On the other hand, when the cluster is much larger than the size of the low-energy spots [Fig. \ref{fig1}(h)], the oscillation becomes smaller. Note that the above picture of the potential energy oscillation is valid for contacts of arbitrary $\delta$. This is verified in Fig. S2 of the Supplemental Material \cite{supplemental}, which shows similar potential-energy oscillations when moir\'e spots are crossing the edges of clusters rotating on surfaces of different $\delta$.

\begin{figure*}[htbp]
  \centering
  \includegraphics[width=1.0\columnwidth]{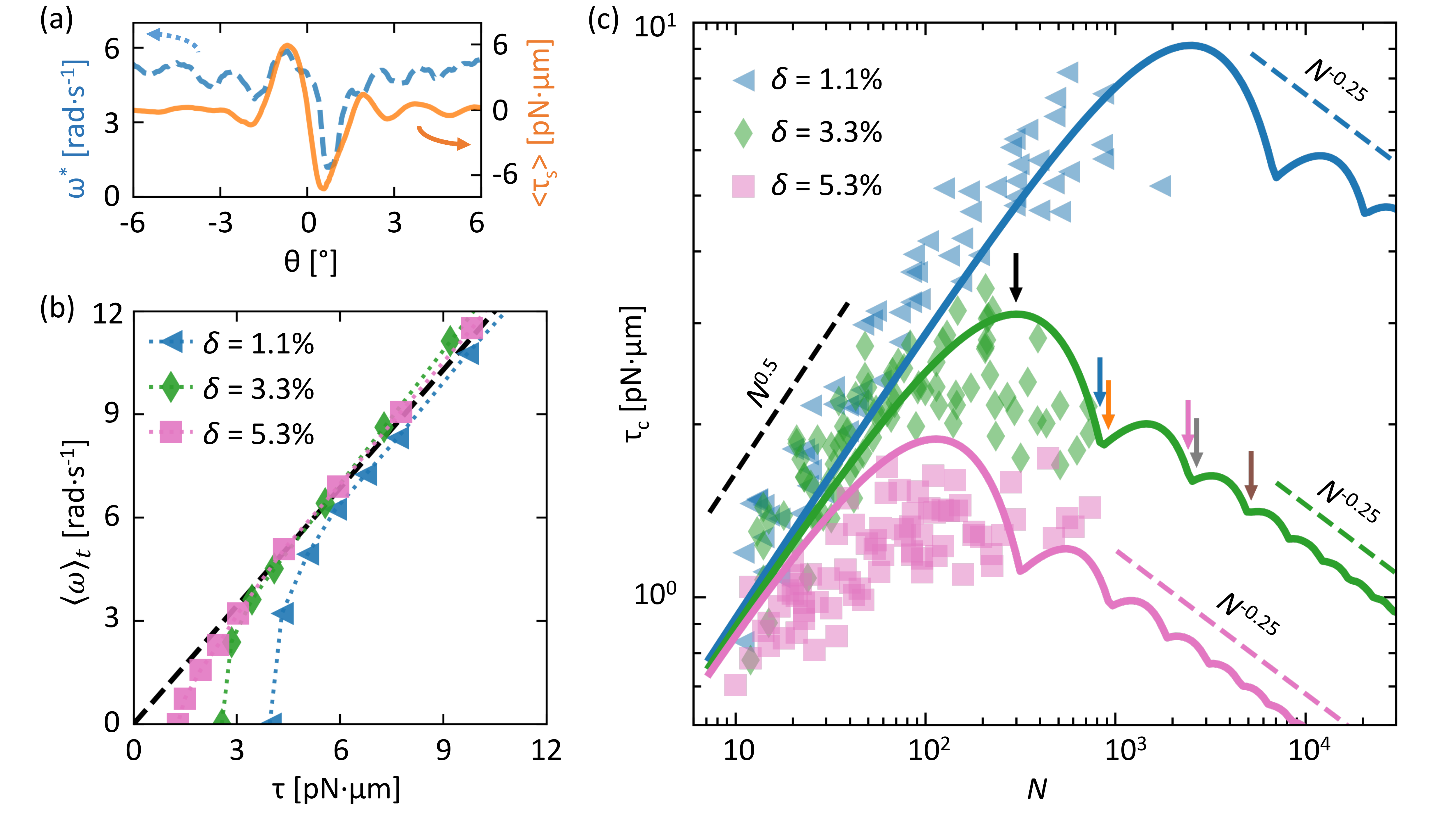}
\caption{Critical depinning torque and its finite-size scaling. (a) Measured instantaneous rotational velocity $\omega^{*}$ and computed average substrate torque $\langle\tau_\mathrm{s}\rangle$ as a function of $\theta$ for the cluster with $N = 1715$ rotating on a $\delta=1.1\%$ surface under applied torque $\tau = 6.68~\mathrm{pN}\cdot\mu\mathrm{m}$. Arrows point at the relevant axis for each curve. At any time instant $\omega^{*}$ is measured over a time interval of 3.3 seconds. (b) The measured average rotational velocity $\langle\omega^{*}\rangle_t$ as a function of the applied torque $\tau$ for three clusters of similar size ($N = 133, 114, 179$) in experiments rotating on $\delta=1.1\%,~3.3\%,~5.3\%$ surfaces respectively. $\langle~\rangle_t$ denotes an average over a time interval where the cluster rotates more than 180 degrees. The data approach the linear relation (dashed line) $\langle\omega^{*}\rangle_t = \tau / (\pi\eta a^3)$ at large $\tau$ (viscosity $\eta = 3.2 \times 10^{-3}~\mathrm{pN}\cdot\mathrm{s}/\mu\mathrm{m}^2$).
On the other hand, the clusters stop rotating when $\tau$ drops below a critical value $\tau_\mathrm{c}$. (c) Data points: measured critical torque $\tau_\mathrm{c}$ as a function of cluster size $N$ on surfaces of different $\delta$. Solid lines: corresponding theoretical curves from Eq. (\ref{eq3}) for circular clusters. Dashed lines indicate scaling relations as discussed in the text. Arrows along the green curve pinpoint cases discussed in Fig. \ref{fig3} below.}
\label{fig2}
\end{figure*}

\subsection{Scaling behaviors of static friction torque}
The above mentioned energy oscillation leads to a torque $\langle\tau_s\rangle=-\partial\langle\epsilon_\mathrm{s}\rangle/\partial \theta$, which we refer to as the substrate torque since it is acting on the cluster by the substrate. In combination with the constant external torque $\tau$ they determine the angular velocity $\omega^{*} = (\tau + \langle\tau_\mathrm{s}\rangle) / (\pi\eta a^3)$. 
This relation is found in good agreement with our experiments which demonstrate an approximate proportionality between $\omega^{*}$ and $\langle\tau_\mathrm{s}\rangle$ in the presence of a constant torque $\tau$ [Fig. \ref{fig2}(a)]. 
To allow for a continuous cluster rotation, the applied torque $\tau$ must exceed a critical value $\tau_\mathrm{c}$ (i.e. the onset of cluster rotation), to satisfy $\omega^{*} \propto \tau + \langle\tau_\mathrm{s}\rangle > 0$ for all $\theta$. 
To determine $\tau_\mathrm{c}$ we have gradually decreased $\tau$ and measured the average rotational velocity $\langle\omega^{*}\rangle_t$ as a function of $\tau$ [see Fig. \ref{fig2}(b)]. 
Note that $\tau_\mathrm{c}$ is smaller for clusters on larger-$\delta$ surface. In general, $\tau_\mathrm{c}$ also depends on the cluster size. This is seen in Fig. \ref{fig2}(c) which shows experimentally measured values of $\tau_\mathrm{c}$ as a function of $N$ for three different $\delta$ (symbols). Despite significant scatter in the data due to different cluster shapes (see Appendix \ref{appendmicromodel}), the following features are observed in our experiments: (i) for nearly matching conditions ($\delta=1.1\%$), $\tau_\mathrm{c} \propto N^{0.5}$ up to $N \sim 1000$. (ii) for slightly larger mismatches ($\delta=3.3\%$), such scaling is satisfied for $N \le \sim100$, and a maximum of $\tau_\mathrm{c}$ is observed around $N\approx200$. (iii) for $\delta=5.3\%$, $\tau_\mathrm{c}$ becomes nearly independent of the cluster size for $N > \sim50$. 
Note that the scaling behaviour of rotational friction torque observed here is very different from that of the translational friction force \cite{ritter2005prb,dietzel2008prl,dietzel2013prl}, not just because rotation and translation involve different degrees of freedom. Moreover, friction torque and friction force require different ways to measure.  

\begin{figure*}[!ht]
  \centering
  \includegraphics[width=1.0\columnwidth]{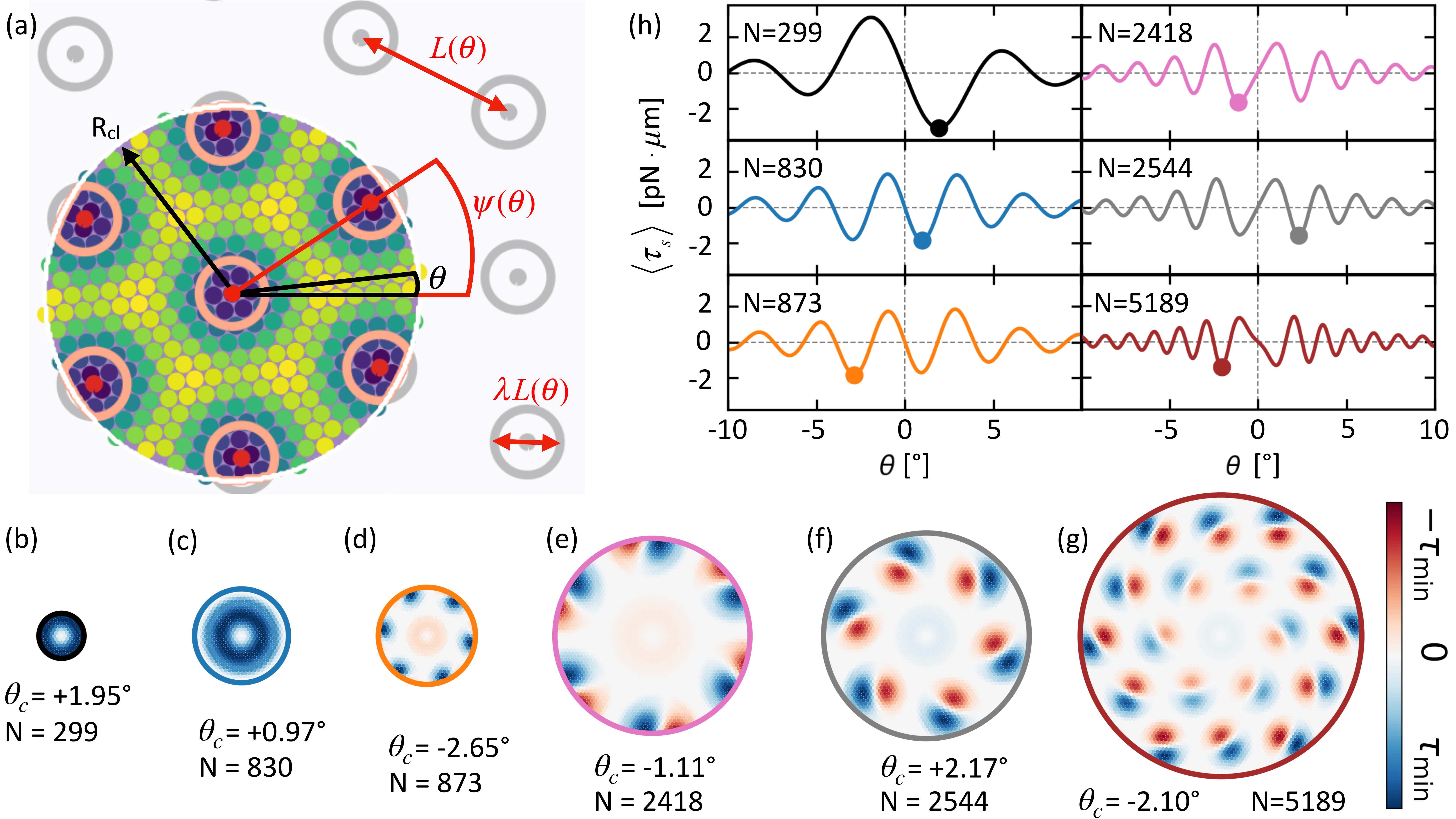}
\caption{Angle dependence of the average substrate torque of different-sized clusters and the corresponding critical configurations. (a) Comparison of the microscopic model and the coarse-grained model. We assume circular-shaped clusters with lattice misalignment angle $\theta$ relative to the substrate lattice. In the microscopic model, discrete particles are considered and their energies [color code as in Fig. \ref{fig1}(e-h)] are summed up to calculate the interaction energy $\langle\epsilon_\mathrm{s}\rangle$. Instead, the coarse-grained model focuses on the periodically-arranged moir\'e spots (red and gray circles). These moir\'e spots are Gaussian energy density profiles of width $\lambda L$ which contribute to $\langle\epsilon_\mathrm{s}\rangle$ if they are covered by the cluster (red). (b-g) Local torque distributions of simulated (with the microscopic model) circular clusters at their depinning angles $\theta = \theta_\mathrm{c}$ (reported in the corresponding panel) on a $\delta=3.3\%$ surface. Individual particles are color coded according to their local substrate torque $\tau_\mathrm{s}$ (Appendix \ref{appendmicromodel}). The $\tau_\mathrm{c}$ values for the six cluster sizes reported here are marked by arrows in Fig. \ref{fig2}(c). (h) $\langle\tau_\mathrm{s}\rangle$ versus $\theta$, as obtained from Eq. (\ref{eq2}) (coarse-grained model) for $\mathbf{r}_\mathrm{cm}=0$ and $b = 4.6~\mu\textrm{m}$ ($\delta=3.3\%$) for the six cluster sizes as shown in (b-g). The filled dots on the curves mark the respective global minima $\theta_c$.}
\label{fig3}
\end{figure*}

\section{Theoretical analysis}
\subsection{The analytical model}
The above experimental findings are well reproduced by numerical simulations of a microscopic model (Appendix \ref{appendmicromodel}) which explicitly considers all particle-surface interactions as in Fig. \ref{fig1}(e-h) and which can be applied to clusters of arbitrary size and shape. In the following, we will demonstrate that the above results are quantitatively reproduced by a much simpler coarse-grained model which allows an analytical formulation of the cluster-surface interaction energy. Within this analytical framework we are able to provide a clear physical understanding how rotational friction depends, e.g., on cluster size and lattice mismatch.
To construct our analytical model, we treat the cluster as a circular disk of radius $R_\mathrm{cl}\propto N^{0.5}$, cutting a finite region of the moir\'e pattern [Fig. \ref{fig3}(a)].
The moir\'e spots are centered on the lattice points of a triangular grid with lattice spacing $L$ and are each described by a Gaussian energy density profile of strength $\epsilon/(\sqrt{3}a^2/2)$ and width $\lambda L$.
The cluster-surface interaction is approximated by integration of all the Gaussian profiles within the cluster area, see Eq. \ref{eqM7} in Appendix \ref{appendanalyticalmodel}.
Due to the interplay between the contacting lattices, a rotation $\theta$ of the cluster results in a rotation $\psi=\psi(\theta)$ of the moir\'e pattern accompanied with a shrinkage or expansion of the lattice spacing $L=L(\theta)$; a cluster translation $\mathbf{r}_\mathrm{cm}$ results in a translation $\mathbf{t}=\mathbf{t}(\theta, \mathbf{r}_\mathrm{cm})$ of the moir\'e pattern \cite{hermann2012jpcm}.
The integration yields an analytic expression for the interaction energy per particle $\langle\epsilon_\mathrm{s}\rangle$ as a function of $\theta$ and $\mathbf{r}_\mathrm{cm}$, i.e. 
\begin{equation}
\begin{split}
\langle \epsilon_\mathrm{s}\rangle(\theta, \mathbf{r}_\mathrm{cm}) =  -\epsilon_\mathrm{eff}\frac{L}{R_\mathrm{cl}} J_1\left(\frac{4\pi}{\sqrt{3}}\frac{R_\mathrm{cl}}{L}\right) \cdot\sum_{n=0}^{5}\cos\left[\frac{4\pi }{\sqrt{3}b} \hat{x}\cdot\mathcal{R}\left(\frac{2n\pi+\pi}{6}\right)\mathbf{r}_\mathrm{cm}\right].
\end{split}
\label{eq1}
\end{equation}
Here $\epsilon_\mathrm{eff}=8\pi^2\lambda^2\epsilon \exp(-8\pi^2\lambda^2/3)$, $J_1()$ the first-order Bessel function of the first kind, $\hat{x}=(1,0)$ a unit vector, and $\mathcal{R}(\varphi)\mathbf{r}_\mathrm{cm}$ rotates the vector $\mathbf{r}_\mathrm{cm}$ by an angle $\varphi$ counterclockwise. Note that rotation and translation contribute separately to $\langle\epsilon_\mathrm{s}\rangle(\theta, \mathbf{r}_\mathrm{cm})$ through the $LJ_1$ and the cosine term, respectively. The $\langle \epsilon_\mathrm{s}\rangle$ calculated with Eq. (\ref{eq1}) shows excellent agreement with that obtained from the microscopic model, see Fig. S3-S5 of the Supplemental Material \cite{supplemental}.

Differentiating Eq. (\ref{eq1}) with respect to $\theta$ we obtain the following expression for the mean substrate torque:
\begin{equation}
  \begin{split}
     \langle\tau_s\rangle = & -\epsilon_\mathrm{eff}\frac{4\pi L^2\sin\theta}{\sqrt{3}ab} J_2\left(\frac{4\pi}{\sqrt{3}}\frac{R_\mathrm{cl}}{L}\right) 
     \cdot\sum_{n=0}^{5}\cos\left[\frac{4\pi }{\sqrt{3}b} \hat{x}\cdot\mathcal{R}\left(\frac{2n\pi+\pi}{6}\right)\mathbf{r}_\mathrm{cm}\right].
  \end{split}
\label{eq2}
\end{equation}
Here $J_2()$ is the second-order Bessel function of the first kind. The critical torque for rotational depinning
\begin{equation}
 \tau_{\mathrm{c}} = -\langle\tau_s\rangle(\theta=\theta_\mathrm{c},\mathbf{r}_\mathrm{cm}=0) 	
 \label{eq3}
\end{equation}
is reported in Fig. \ref{fig2}(c) as a function of cluster size $N$ (solid lines) for $\delta=1.1\%,~3.3\%,~5.3\%$, where $\theta_\mathrm{c}$ denotes the angle where $\langle\tau_\mathrm{s}\rangle$ reaches its minimum for $\mathbf{r}_\mathrm{cm}=0$ (i.e. pure rotation).

\subsection{Theoretical understandings of the cluster-size-dependence of the critical torque}
The above analytical results confirm the experimentally observed $N^{0.5}$ scaling at small $N$. In addition, Eq. (\ref{eq3}) predicts a strict $N^{0.5}$ scaling at all cluster size $N$ for the $\delta=0$ contact (see Appendix \ref{appendanalyticalmodel}). Such scaling results from the coherent summation of all local substrate torques $\tau_\mathrm{s}$ inside the cluster, which applies when the cluster size is smaller than a single moir\'e spot. This is exemplarily shown in Fig. \ref{fig3}(b,c) where the local torques of two small clusters at their depinning angles $\theta = \theta_\mathrm{c}$ are obtained from the microscopic model. Even though they differ in amplitude, all local torques have the same sign upon cluster depinning which rationalizes a coherent summation. Equation (\ref{eq3}) shows that critical torques for different $\delta$ in Fig. \ref{fig2}(c) can overlap into a single universal curve applicable to all mismatches (see Appendix \ref{appendanalyticalmodel}). Interestingly, Eq. (\ref{eq3}) also suggests an oscillatory behavior of $\tau_\mathrm{c}$ as a function of $N$ for all contacts with $\delta>0$. To rationalize such behavior, Fig. \ref{fig3}(d-g) illustrates the local torque distributions for four simulated clusters of increasing size at their corresponding $\theta_\mathrm{c}$. Summation of such local torques yields an oscillation the of $\tau_\mathrm{c}$-$N$ relation (see Appendix \ref{appendmicromodel}), which is fully captured by Eq. (\ref{eq3}). Opposed to the small clusters of Fig. \ref{fig3}(b,c), torques of both signs are observed in Fig. \ref{fig3}(d-g), where cluster sizes are large enough to accommodate a first and second ring of moir\'e spots. This leads to drastic changes in the local torque distribution for clusters near certain sizes when a new moir\'e ring enters their edge [e.g. Fig. \ref{fig3}(c) vs. Fig. \ref{fig3}(d) and Fig. \ref{fig3}(e) vs. Fig. \ref{fig3}(f)], and eventually leads to the observed oscillatory behavior of $\tau_\mathrm{c}(N)$ in Fig. \ref{fig2}(c). At the same time, $\theta_\mathrm{c}$ changes between positive and negative values at these sizes. By comparison, Fig. \ref{fig3}(h) reports the substrate torque per particle $\langle\tau_{\textrm{s}}\rangle$ obtained from Eq. (\ref{eq2}) as a function of the cluster’s orientation $\theta_\mathrm{c}$ for six circular clusters with the same size as shown in Fig. \ref{fig3}(b-g). The absolute minima denote the corresponding $\theta_\mathrm{c}$ which reveal similar sign changes as a function of size. Such sign changes are also observed in experiments for clusters rotating on $\delta=1.1\%,~3.3\%,~5.3\%$ surfaces (see Fig. S6 of the Supplemental Material \cite{supplemental}). 

\begin{figure*}[!htbp]
  \centering
  \includegraphics[width=1.0\columnwidth]{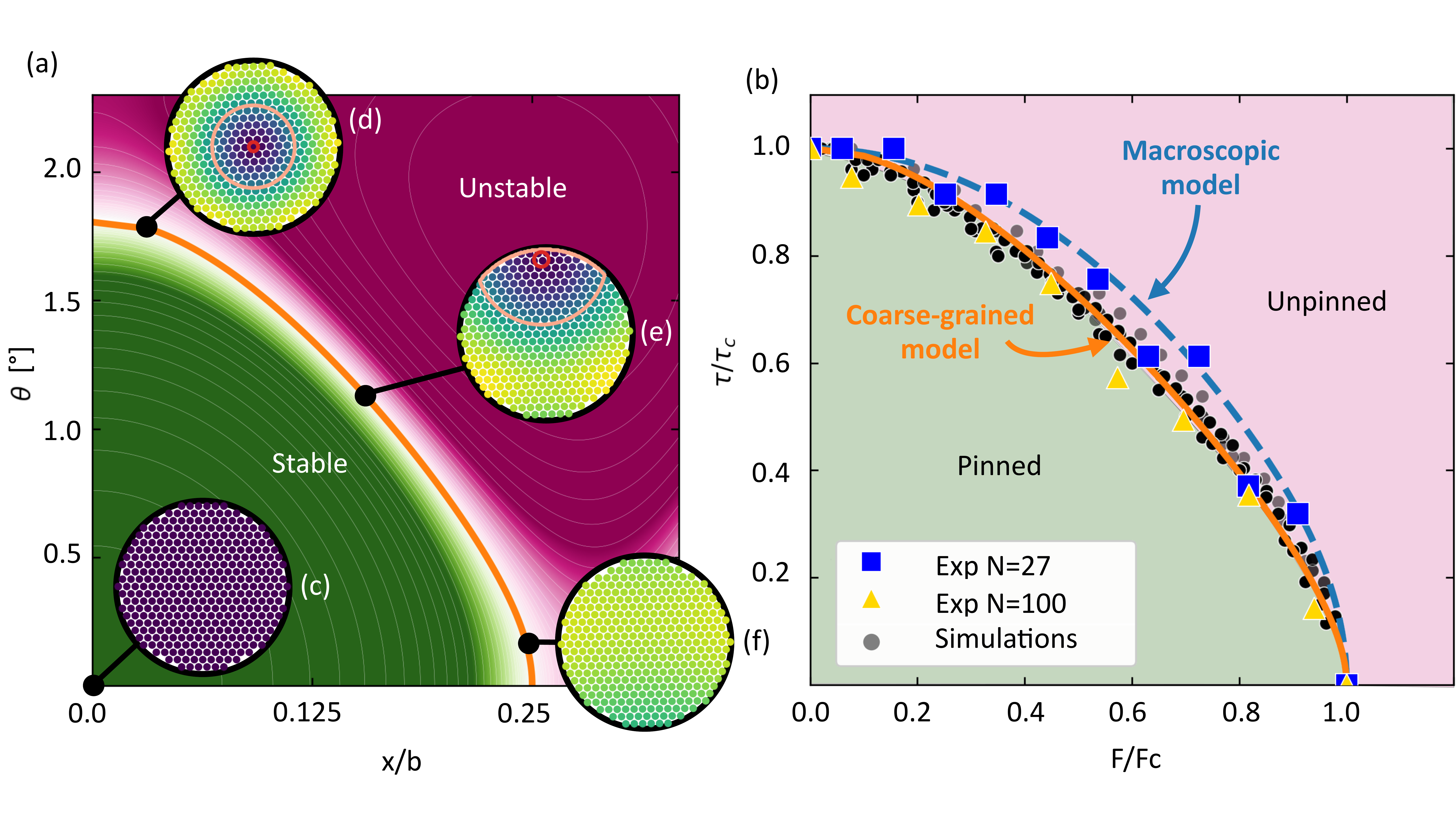}
\caption{Rotation-translation depinning boundary and its mechanism.  (a) The calculated $\det \mathcal{H}$ (color code) as a function the center-of-mass rotation $\theta$ and translation $x$, for a cluster with $N = 363$ ($R_\mathrm{cl}=44.5~\mu\mathrm{m}$) on a $\delta=0$ surface. Green and pink regions correspond to $\det \mathcal{H}>0$ and $\det \mathcal{H}<0$ respectively. The orange line marks the $\det \mathcal{H}=0$ depinning boundary. (b) The measured $(\tau, F)_\mathrm{c}$ pairs for two experimental clusters with $N = 27$ and $N = 100$ on a $\delta=1.1\%$ surface (squares and triangles), and for simulated clusters with a range of sizes (from $N = 15$ to 5000) and shapes (circular, hexagonal) on a $\delta=0$ surface (dots). The force in experiments is achieved by tilting the sample at an angle $\alpha$, which applies gravitational force $F = mg\sin\alpha$ to colloidal particles of buoyant weight $mg$ \cite{cao2019np,cao2021pre}. In both experiments and simulations, to measure the $(\tau, F)_\mathrm{c}$ pairs, we apply a force $F$ in the $x$ direction and then gradually increase the torque $\tau$ until the cluster depins. All simulation points collapse near the stability-boundary line (solid), obtained within the coarse-grained model. (c-f) Snapshots of the $N=363$ simulated circular cluster shown in (a) at four different coordinates $(\theta,x)$ indicated by the corresponding linked dot. The colloids are color coded according to the substrate potential as in Fig. \ref{fig1}(e-h), showing the moir\'e pattern evolution when the cluster coordinates change.}
\label{fig4}
\end{figure*}

Remarkably, since the oscillation amplitude of the integer-order Bessel functions $J_n(z)$ decays as $1/\sqrt{z}$, Eq. (\ref{eq3}) predicts an asymptotic behaviour $\tau_\mathrm{c} \propto N^{-0.25}$ at large $N$ (see Appendix \ref{appendanalyticalmodel}), which is also reproduced by simulations of circular clusters (see Appendix \ref{appendmicromodel}). 
Interestingly, this scaling leads to a sublinear relation of the total static torque $\Gamma_\mathrm{c} = N\tau_\mathrm{c} \propto N^{0.75}$ which suggests a superlow-static-torque state. 
Although it is mathematically different from a superlubric state where the contacting surfaces can depin without resistance \cite{shinjo1993ss}, such scaling enables extremely small rotational friction per unit area for sufficiently large contacts in nanomechanical components. 
Note that such a state only occurs for $\delta>0$ and circular contacts. For non-circular contacts, such as hexagons, squares and triangles, a $\tau_\mathrm{c} \propto N^{0}$ scaling is observed at large $N$ in numerical simulations and in experiments, see Appendix \ref{appendmicromodel}. The observed $\tau_\mathrm{c} \propto N^{0.5}$, $\tau_\mathrm{c} \propto N^{0}$ and $\tau_\mathrm{c} \propto N^{-0.25}$ scaling relations agree well with an extension of an empirical scaling law obtained for translational friction (see Appendix \ref{appendscaling}). Our findings regarding the size scaling are summarized in Table S1 of the Supplemental Material \cite{supplemental}.

\subsection{Translational and orientational friction coupling}
To study the interplay of translational and rotational depinning, we consider the analytic expression of the generalised enthalpy $A(\theta, \mathbf{r}_\mathrm{cm}) = \langle\epsilon_\mathrm{s}\rangle(\theta, \mathbf{r}_\mathrm{cm}) - \theta\tau - \mathbf{r}_\mathrm{cm}\cdot\mathbf{F}$, which determines the equilibrium position of the cluster in presence of an external driving torque $\tau$ and force $\mathbf{F}$. Instability (i.e. depinning) occurs when the determinant of the Hessian matrix $\det \mathcal{H}$ of the second derivatives of $A(\theta, \mathbf{r}_\mathrm{cm})$ turns negative (see Appendix \ref{appendanalyticalmodel} for details). For simplicity we only consider translations along the $x$ direction, i.e. $\mathbf{r}_\mathrm{cm}=(x,0)$. The calculated $\det \mathcal{H}$ as a function of $\theta$ and $x$ is reported in Fig. \ref{fig4}(a) for a cluster of $N=363$ and $\delta=0$. A boundary between the stable (green) and unstable (pink) region is indicated by $\det \mathcal{H} = 0$ (orange solid line), which marks the critical displacement $(\theta,x)_\mathrm{c}$. Figure \ref{fig4}(b) reports the critical drive $(\tau,F)_\mathrm{c}$ corresponding to $(\theta,x)_\mathrm{c}$, which is in excellent agreement with the experimental and simulated results obtained for clusters of very different sizes and shapes. Note that the $(\tau,F)_\mathrm{c}$ line (solid) obtained for circular colloidal clusters on perfect crystalline surfaces systematically falls below that obtained for a macroscopic disc in contact with a uniform surface (dashed curve) \cite{dahmen2005pre}. The difference originates from a fundamentally distinct depinning mechanism: compared to the spatially uniform depinning of rigid macroscopic contacts \cite{dahmen2005pre}, the depinning of our colloidal cluster depends on the rotation- and translation-induced moir\'e-pattern evolution as shown in Fig. \ref{fig4}(c-f) and Video 4. In the case when only torque (force) is involved, the cluster leaves the initial perfectly commensurate configuration [Fig. \ref{fig4}(c)] and moves along the torque-driven (force-driven) direction. The moir\'e pattern, which starts to develop, shrinks (translates) uniformly as shown in Fig. \ref{fig4}(d) [Fig. \ref{fig4}(f)] until the cluster depins. In situations where both torque and force are involved, the moir\'e spot becomes located at one side of the cluster as shown in Fig. \ref{fig4}(e). In this case, depinning is favourably triggered by the emergence, at the opposite side of the cluster, of a locally incommensurate weak-pinning region of the moir\'e pattern which then leads to depinning of the entire system. This non-uniform depinning mechanism dominates as long as the moir\'e spots are not symmetrically distributed in the cluster. This mechanism, remarkably observed in a $\delta=0$ contact, becomes particularly relevant for the $\delta>0$ contacts, where the depinning is determined by the preformed moir\'e spots at the cluster’s edge, which leads to further deviation of the $(\tau,F)_\mathrm{c}$ line from that of the $\delta=0$ case, as shown by the numerical simulations in Fig. S7 of the Supplemental Material \cite{supplemental}.

\section{Discussions}
The complex depinning of torque- and force-driven colloidal clusters on crystalline surfaces as demonstrated here should be of immediate relevance for nano-manipulation experiments where e.g. atomic-force microscopes often induce not only forces but additional torques which drastically affect the depinning of nanoparticles and their translational friction \cite{filippov2008prl,deWijn2011epl}. 
Similar to macroscopic scales where e.g. circular-shaped clutches or end bearings are used to achieve smooth friction forces, the super-low static rotational friction state found in our work suggests that circular contacts also provide the ideal contact geometry at microscopic scales. 
This may be useful for the design of atomic actuators and nano-electro-mechanical-devices where low rotational friction is desired. 
Finally, the complex moir\'e pattern evolution upon cluster rotation may find use in the area of twistronics where angle-dependent variations of the electronic properties between atomically flat layers are exploited for applications \cite{ribeiro2018sci}. 

\section*{Acknowledgement}
X.C. and C.B. acknowledge financial support from the Alexander von Humboldt Foundation and the CRC 1214 (Deutsche Forschungsgemeinschaft). E.T. acknowledges support by ERC ULTRADISS Contract No. 834402. N.M., A.V.  and A.S. acknowledge support by the Italian Ministry of University and Research through PRIN UTFROM N. 20178PZCB5. A.S. and A.V. acknowledge support by the European Union’s Horizon 2020 research and innovation programme under grant agreement No. 899285. We would like to thank Dieter Barth, Florian Zaunberger and Thomas Trenker for their technical support in realizing the rotating magnetic field.

\appendix

\section{Sample preparation, cluster formation and image analysis.}
\label{appendsample}
We use superparamagnetic colloidal spheres (Dynabeads M-450) with diameter $a = 4.45~\mu\mathrm{m}$ which are dispersed in an aqueous sodium dodecyl sulfate (SDS) solution at 90\% of the critical micellar concentration. The concentration of colloidal particles is about $2\times10^8/\mathrm{mL}$. As flocculating agent we use an aqueous solution of 0.02-weight-percent polyacrylamide (PAAm) with molecular weight 18,000,000 a.m.u. Patterned substrates are prepared by first spin coating a glass surface with a thin layer ($\sim 80~\mathrm{nm}$) of SU8 photoresist. Afterwards it is exposed to ultraviolet light through a photo mask that contains the corresponding surface pattern. After development a patterned area on the substrate with dimensions $16~\mathrm{mm} \times 40~ \mathrm{mm}$ is obtained. To achieve a closed sample cell, we first apply two parafilm spacers with $\sim 150~\mu\mathrm{m}$ thickness at two opposite sides of the patterned area and glue a cover slide on top of it. Then, a mixture of the suspension containing $5~\mu L$ of the colloid-SDS solution and $100~\mu L$ of the flocculant solution is injected at the open ends and sealed afterwards with epoxy glue. Since the colloidal spheres are heavier than water (buoyant weight $mg = 286~ \mathrm{fN}$), they sediment towards the bottom substrate of the sample cell. Due to the flocculation effect, colloidal spheres stick tightly together once they come into contact (e.g. via diffusion). To accelerate the formation of large colloidal clusters, we tilt the sample at $20^{\circ}$, so that emerging colloidal clusters drift over the entire substrate. During this process they grow in size by collecting more and more colloids (see Fig. S8 of the Supplemental Material \cite{supplemental}). This process yields 2D crystalline clusters with a broad distribution of size (up to $N = 1700$ particles) and shape. 
As shown in Fig. S9 of the Supplemental Material \cite{supplemental}, these clusters have an extremely small nearest-neighbour bond length fluctuation (0.34\%) during their rotation on the periodic surfaces. This leads to a critical size of about $N_\mathrm{c}=(1/0.0034)^2\approx90000$ particles, below which the cluster's elasticity effect can be negligible.
Like in previous work \cite{cao2019np}, we obtain the positions of the colloidal particles and those of the substrate wells simultaneously by using computer microscopy as shown in Fig. S10 of the Supplemental Material \cite{supplemental}. This allows us to know the positions of the colloidal particles relative to the substrate wells. 

One of the advantages of our colloidal model system is that we can measure the cluster’s orientation in a very precise manner. Specifically this is done by measuring the average orientation of all nearest-neighbor bonds in one lattice direction. For a single bond, the error of its orientation is roughly 0.5 $^{\circ}$, which is estimated from the uncertainty in the particle position (approximately 40 nm or 1/3 pixel size) divided by the interparticle spacing (4.45 $\mu$m). The more particles in the cluster, the more nearest-neighbor bonds are involved in the calculation of the cluster orientation, and the more precisely the cluster angle is measured. A rough estimation of the precision of the orientation of a cluster composed of $N$ particles is $0.5^{\circ}/\sqrt{2N}$. Here the factor $2N$ comes from the fact that each particle has, on average, two bonds along a certain lattice direction. Taking, for example, a cluster of 50 particles, the precision of the angle is roughly $0.05^{\circ}$. For the cluster shown in Fig. \ref{fig1}(e-h), it contains $N = 1715$ particles, which yields a precision of about $0.008^{\circ}$. Note that these estimations assume independent and normal-distributed uncertainties in individual bond angle measurement, the real precisions will even be better due to the existence of correlations inherent from sharing of bonding particles.

\section{Viscous rotation, magnetic torque formulation and calibration.}
\label{appendcalibration}
The viscous torque of a rigid colloid cluster rotating with angular velocity $\omega$ in a liquid suspension can be expressed as \cite{ranzoni2010lc} $\Gamma_\mathrm{v} = \sum_i (\tau_i + r_iF_i)$. Here $\tau_i = 8\pi\eta(a/2)^3\omega$ is the viscous torque of a single colloidal sphere rotating at angular velocity $\omega$ around its center of mass, $F_i = 6\pi\eta(a/2)\omega r_i$ the viscous force acting on the colloid when moving at speed $\omega r_i$ in the suspension, $r_i$ the distance of particle $i$ to the axis of rotation (the cluster’s center-of-mass position for our 2D clusters), $\eta$ the solvent’s viscosity, and $a = 4.45~\mu\mathrm{m}$ the colloidal diameter. With the above quantities this yields $\Gamma_\mathrm{v} = 8N\pi\eta(a/2)^3\omega + 6\pi\eta(a/2)\omega\sum_i r_i^2$. With the shape-dependent factor $I = \sum_i 3r_i^2 / (Na^2)$, this finally leads to
\begin{equation}
\Gamma_\mathrm{v} = N\pi\eta a^3(1+I)\omega.    
\label{eqM1}
\end{equation}
The factor $I$ characterizes the cluster shape. For a one-dimensional periodic chain of particles, $I(N) \propto N^2$. For a two-dimensional cluster $I(N) \propto N$. In our experiments, we have chosen compact two-dimensional clusters where a linear relation $I(N) \propto N$ is revealed (see Fig. S11 of the Supplemental Material \cite{supplemental}). 

A rotating magnetic field $\mathbf{H}$ induces a rotating magnetization $\mathbf{M}_i$ in each colloidal particle $i$ within a cluster. Due to a small phase lag in $\mathbf{M}_i$, a torque $\Gamma = |\sum_i \mathbf{M}_i \times \mathbf{H}|$ acts on the entire cluster. Since the magnetization of particles within a cluster is slightly screened by their neighbours, the total magnetization $\sum_i\mathbf{M}_i$, and thus $\Gamma$, depends on the size and shape of the cluster. Considering that our colloidal spheres have a rather uniform size (polydispersity $<$ 5\%), $\Gamma$ can be calculated by classifying the colloidal particles within the cluster as either bulk particles (coordination number = 6) or edge particles (coordination number $<$ 6). Therefore, $N = N_\mathrm{bulk} + N_\mathrm{edge}$ and $\sum_i\mathbf{M}_i = N_\mathrm{bulk}\mathbf{M}_\mathrm{bulk} + N_\mathrm{edge}\mathbf{M}_\mathrm{edge}$. Here $\mathbf{M}_\mathrm{bulk}$ is the magnetization of the bulk particles and $\mathbf{M}_\mathrm{edge}$ is the magnetization of the edge particles. We assume $\mathbf{M}_\mathrm{edge} = l\mathbf{M}_\mathrm{bulk}$ where $l$ is a parameter that describes the edge-particle magnetization relative to that of the bulk particle. For simplicity we treat $l$ as a constant in our experiments. Accordingly, the total magnetization $\sum_i\mathbf{M}_i = \rho_\mathrm{M}N\mathbf{M}_\mathrm{bulk}$, with $\rho_\mathrm{M} = 1 - (N_\mathrm{edge}/N)(1-l)$ describing the influence of the edge particles. This leads to $\Gamma = \rho_\mathrm{M}N|\mathbf{M}_\mathrm{bulk}\times \mathbf{H}|$. Considering a linear response $|\mathbf{M}_\mathrm{bulk}| \propto H = |\mathbf{H}|$, we finally obtain
\begin{equation}
\Gamma =  \rho_\mathrm{M}NkH^2,
\label{eqM2}
\end{equation}
where $k$ depends on the magnetic susceptibility of the colloids as well as the misalignment angle between $\mathbf{M}_\mathrm{bulk}$ and $\mathbf{H}$. The balance between magnetic and viscous torques gives $N\pi\eta a^3(1+I)\omega = \rho_\mathrm{M}NkH^2$, or 
\begin{equation}
\omega = \omega_0 \rho_\mathrm{M}/(1 + I),
\label{eqM3}
\end{equation}
where $\omega_0 = kH^2 / (\pi\eta a^3)$. The Video 1 and Fig. S1 of the Supplemental Material \cite{supplemental} clearly shows the smooth rotation of two colloidal clusters on a flat, i.e. unpatterned, substrate when the rotating magnetic field is switched on. According to Eq. (\ref{eqM3}), $\omega$ scales with the square of $H$ which is verified in our experiments for clusters of various sizes and shapes (Fig. S12 of the Supplemental Material \cite{supplemental}). This scaling also demonstrates that $k$ does not depend on the magnetic-field amplitude and rationalizes that it can be considered to be constant in our experiments.

Equation (\ref{eqM2}) allows us to calculate the value of $\Gamma$ for every cluster in our experiments at any given magnetic field $H$, once we know the parameters $k$ and $l$. These two parameters can be fitted from Eq. (\ref{eqM3}) as we measure the value of $\omega$ for $\sim100$ clusters of different size and shape at fixed $H = 244~\mathrm{A/m}$ in a colloidal sample with $\eta = 4.3\times10^{-3}~\mathrm{pN}\cdot\mathrm{s}/\mu\mathrm{m}^2$. The fitting is done by minimising a cost function $C(\omega_\mathrm{cal})= \sum_i(\omega_\mathrm{cal} - \omega_i)^2$, where $\omega_i$ is the measured rotational velocity of a cluster $i$ and $\omega_\mathrm{cal}$ is the corresponding calculated results from Eq. (\ref{eqM3}). The value of $k$ and $l$ that best fits to our experimental measurement is $k = 7.06\times10^{-5}~\mathrm{pN}\cdot\mu\mathrm{m} / \mathrm{(A/m)}^2$ and $l = 0.33$. To demonstrate the fitting, in Fig. S13(a) and (b) of the Supplemental Material \cite{supplemental} we plot the measured $\omega$ as a function of $\rho_\mathrm{M}/(1 + I)$ for the $\sim100$ experimental clusters for $l = 1$ and $l = 0.33$ respectively. According to Eq. (\ref{eqM3}), $\omega$ scales linearly with $\rho_\mathrm{M}/(1 + I)$.  This linear relation is not fulfilled as in Fig. S13(a) of the Supplemental Material \cite{supplemental} when we choose $l = 1$. In contrast, the linear relation is clearly revealed in Fig. S13(b) of the Supplemental Material \cite{supplemental} when we choose $l = 0.33$. The slope of the linear relation is $\omega_0 = 3.54~ \mathrm{rad/s}$, which gives $k = \pi\eta a^3\omega_0 / H^2 = 7.06\times10^{-5}~\mathrm{pN}\cdot\mu\mathrm{m} / \mathrm{(A/m)}^2$.   

\section{Particle-substrate interactions and numerical simulation of the microscopic model.}
\label{appendmicromodel}
To calculate the potential energy $\epsilon_\mathrm{s}=V_\mathrm{well}(\delta r)$ of a colloidal particle placed at a distance $\delta r$ from the center of the nearest potential well, we use the formula:
\begin{equation}
\begin{split}
V_\mathrm{well}(\delta r) &= -\epsilon e^{\frac{\delta r^2}{2 w^2}}, ~~\delta r \le r_\mathrm{m}\\
V_\mathrm{well}(\delta r) &= -\epsilon e^{\frac{\delta r^2}{2 w^2}}f_\mathrm{s}(\rho_\mathrm{r}),~~  r_\mathrm{m} < \delta r < r_\mathrm{M}\\
V_\mathrm{well}(\delta r) &= 0,~~ \delta r \ge r_\mathrm{M}, 								
\end{split}
\label{eqM4}
\end{equation}
with $\rho_\mathrm{r}=(\delta r - r_\mathrm{m})/(r_\mathrm{M}-r_\mathrm{m})$, $r_\mathrm{m}=1.5~ \mu\mathrm{m}$, $r_\mathrm{M}=2.15~ \mu\mathrm{m}$, $\epsilon=270~ \mathrm{zJ} = 66.34 k_\mathrm{B}T$, $w=0.7~\mu\mathrm{m}$, and the function $f_\mathrm{s}(\rho_\mathrm{r}) = 1-10 \rho_\mathrm{r}^3+15\rho_\mathrm{r}^4-6\rho_\mathrm{r}^5$ provides a smooth cutoff to the Gaussian profile which prevents energy cusps and force discontinuities. The parameters $r_\mathrm{m}$, $r_\mathrm{M}$, $w$ and $\epsilon$ are chosen such that Eq. \ref{eqM4} closely resembles the potential profile of a colloidal sphere on the topographic surfaces as shown in Fig. S14 of the Supplemental Material \cite{supplemental}. Given the Eq. (\ref{eqM4}), the  potential energy per particle $\langle\epsilon_\mathrm{s}\rangle$  as calculated in Fig. \ref{fig1}(d) is then the summation of the $V_\mathrm{well}(\delta r)$ for all particles in the cluster divided by $N$. Similarly, the substrate torque $\tau_\mathrm{s}$ of a colloidal particle can be expressed as $\mathbf{\tau}_\mathrm{s} = (\mathbf{r}-\mathbf{r}_\mathrm{cm}) \times \mathbf{\nabla} V_\mathrm{well}(\delta r)$, where $\mathbf{r}-\mathbf{r}_\mathrm{cm}$ is the position of the colloidal particle relative to the center of mass of the colloidal cluster. The $\langle\tau_\mathrm{s}\rangle$ reported in Fig. \ref{fig2}(a) is averaged over all the particles in the cluster.

\begin{figure}
  \centering
  \includegraphics[width=1\columnwidth]{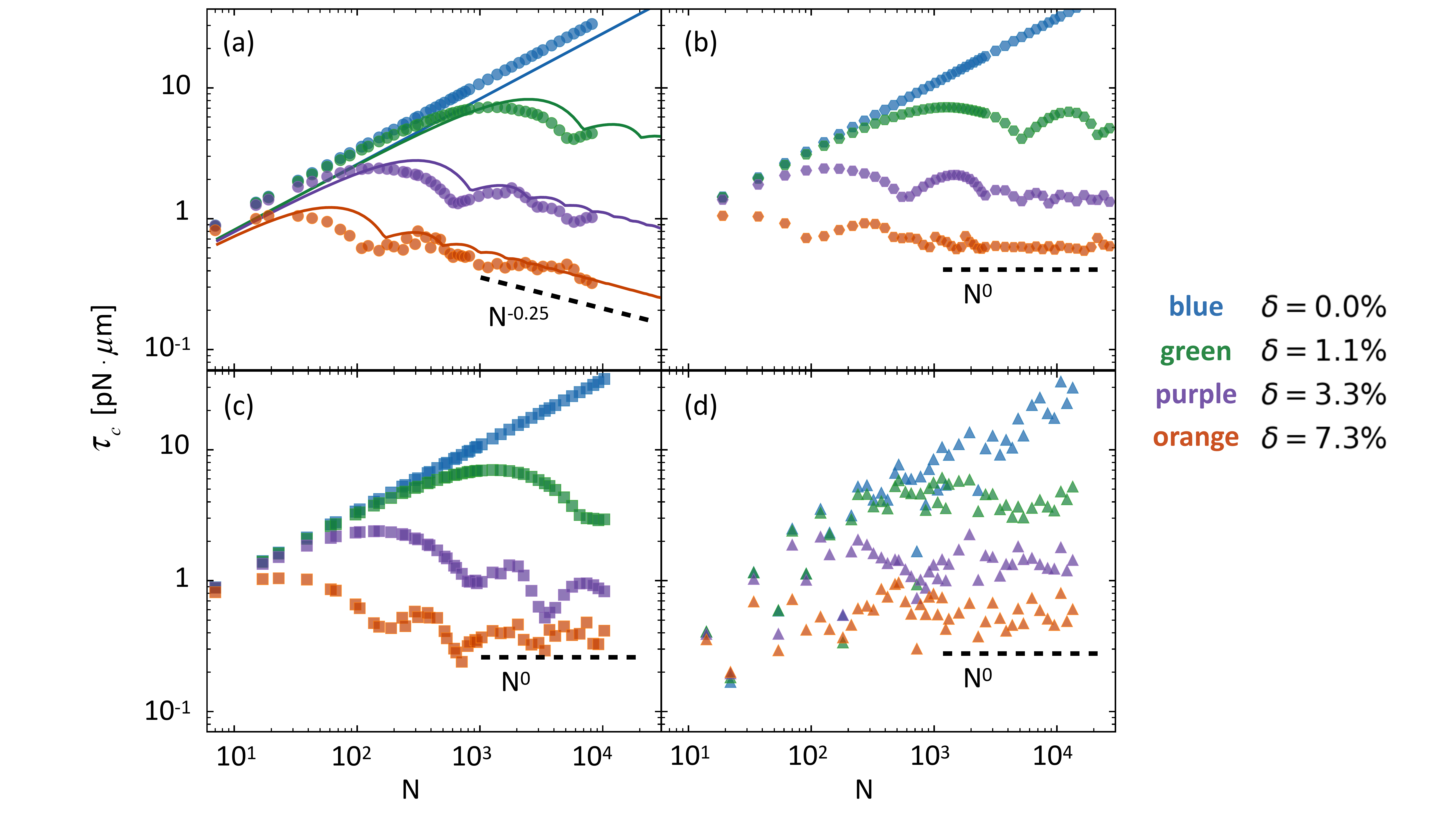}
\caption{The static torque $\tau_\mathrm{c}$ as a function of cluster size $N$ evaluated numerically from the microscopic model (data points) at different mismatch ratio $\delta$. Panels (a), (b), (c), (d) correspond to circular-, hexagonal-, square-, and triangular-shaped clusters respectively. In each panel, the four sets of data (from above to below) corresponds to $\delta=0.0\%,~1.1\%,~3.3\%,~7.3\%$ respectively (also specified by different colors as reported in the legend). A scaling $\tau_\mathrm{c} \propto N^{0}$ is observed at large $N$ for all shapes except for the circular clusters in (a) where a $\tau_\mathrm{c} \propto N^{-0.25}$ relation is observed, in agreement with the corresponding curves of the coarse-grained model (solid lines).}
\label{fig5}
\end{figure}

\begin{figure*}
  \centering
  \includegraphics[width=1\columnwidth]{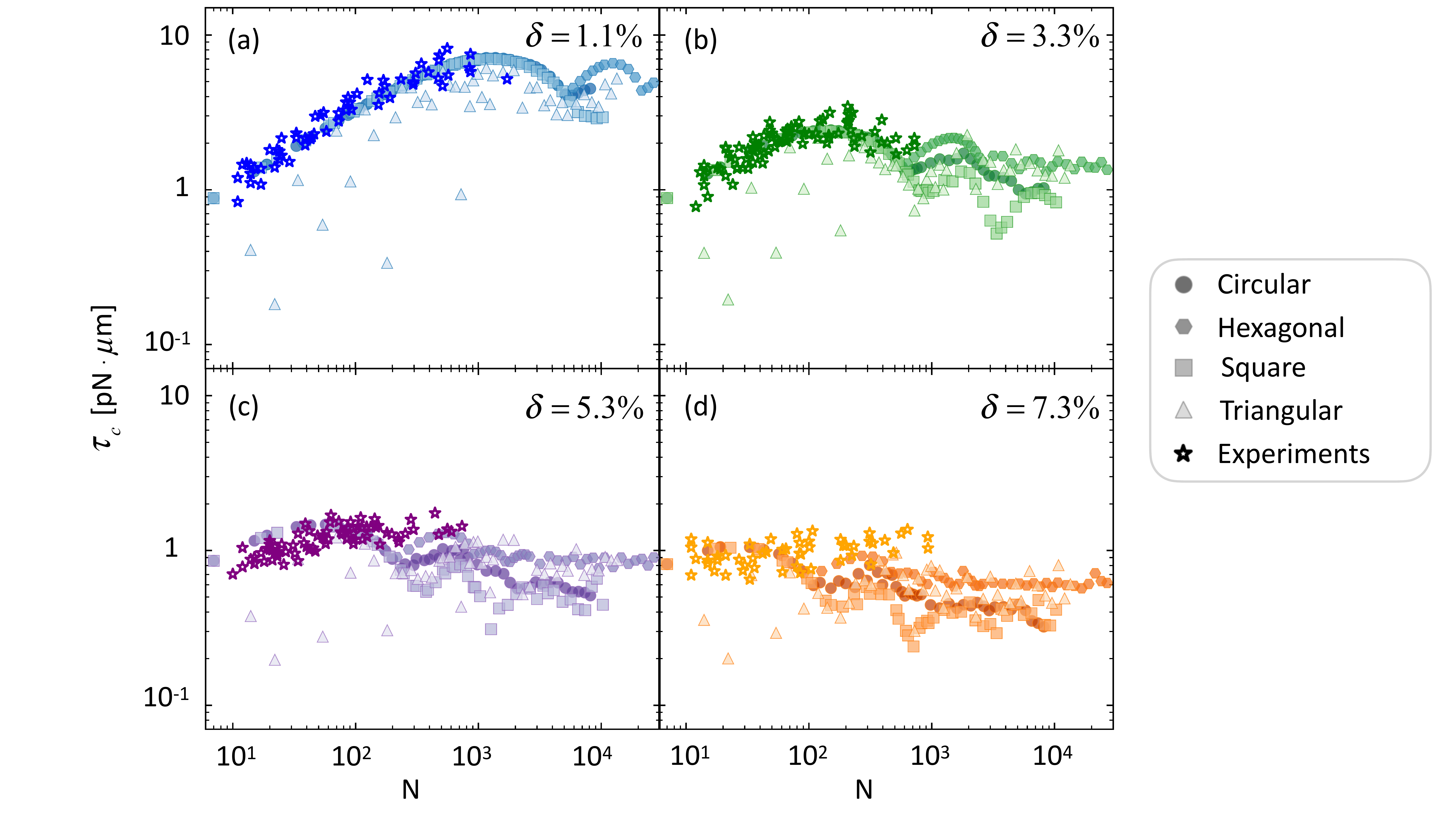}
\caption{The static torque $\tau_\mathrm{c}$ as a function of cluster size $N$ calculated numerically from the microscopic model for circular-, hexagonal-, square-, and triangular-shaped clusters (filled data points) and measured from experiments for irregular-shaped clusters (empty stars). Panels (a), (b), (c), (d) correspond to results of different lattice mismatches $\delta=1.1\%$, $3.3\%$, $5.3\%$ and $7.3\%$ respectively.}
\label{fig6}
\end{figure*}

In simulation, we describe a cluster of colloids as a rigid body with particle positions $\mathbf{r}_i = i_1 \mathcal{R}(\theta)\mathbf{a}_1 + i_2 \mathcal{R}(\theta)\mathbf{a}_2 + \mathbf{r}_\mathrm{cm}$, where $\mathbf{a}_1=(a , 0)$, $\mathbf{a}_2 = (-a/2, \sqrt{3}a/2)$, $\mathcal{R}(\theta)$ the two-dimensional rotation matrix, $\mathbf{r}_\mathrm{cm} = \sum_i \mathbf{r}_i/N$ the center of mass (CM) of the cluster, and the set of integer pairs $i=(i_1,i_2)$ defines the shape and size of the cluster. The shape of the cluster is chosen so that its CM coincides with a particle. The numerically calculated $\tau_\mathrm{c}$-$N$ relation for such clusters of different shape and on substrate of different mismatch ratio $\delta$ is shown in Fig. \ref{fig5}. A comparison of these $\tau_\mathrm{c}$-$N$ relation with those measured in experiments is shown in Fig. \ref{fig6}. The numerical results in Fig. \ref{fig5} and Fig. \ref{fig6} show that clusters of different shapes can have very different values of $\tau_\mathrm{c}$ even they have the same cluster size. This accounts for the large dispersion of experimental data in Fig. \ref{fig2}(c).  

Note that the rigid-contact assumption in our experiments is also well established in various real nanoscale 2D systems over a wide range of sample sizes. For example, an elastic critical length is defined in \cite{sharp2016prb}, below which the dislocation induced elasticity will be negligible at the contact interface. Using the experimental data in \cite{li2010pbcm,liu2012nl,liao2022nmat} and \cite{ma2015prl}, the elastic critical length for MoS$_2$/Graphene heterostructure and double-walled carbon nanotube is calculated to be on the order of millimeters and centimeters respectively, which are already far larger than the contact sizes in most of the relevant experiments.

To simulate the cluster depinning in the presence of external torque and force $(\tau, \mathbf{F})$ as in Fig. \ref{fig4}(b), we assume an overdamped dynamics of the rigid cluster and integrate the first-order Langevin equations of motion
\begin{equation}
\gamma_\mathrm{r} \dot{\theta} = -\sum_i (\mathbf{r}_i - \mathbf{r}_\mathrm{cm}) \times \nabla V_\mathrm{well}(\delta r_i) + \tau + \sqrt{2 T \gamma_r} \eta.
\label{eqM5}
\end{equation}

\begin{equation}
\gamma_\mathrm{t} \dot{\mathbf{r}}_\mathrm{cm} = -\sum_i \nabla V_\mathrm{well}(\delta r_i) + \mathbf{F} + \sqrt{2 T \gamma_t} \mathbf{\eta}.
\label{eqM6}
\end{equation}
The equations are integrated with a time step $dt=5 \cdot 10^{-4}$ ms. The effective rotational and translational viscous-friction coefficients are defined as $\gamma_r = \gamma \sum_i r_i^2$ and $\gamma_t=\gamma N$, respectively, with $\gamma = 1 $ fKg/ms. Since we solve first-order equations, the value of the $\gamma$ does not affect the dynamic behaviour of the system.  $\eta$ is an uncorrelated Gaussian random variable of unit variance.

\section{The analytic coarse-grained model.}
\label{appendanalyticalmodel}
To obtain an analytical form for the cluster-substrate interaction energy and torque, we resort to a coarse-grained model. Consider a cluster translation $\mathbf{r}_\mathrm{cm} = r_{\mathrm{cm}}\mathcal{R}(\theta_\mathrm{d}) \hat{x}$ along direction $\theta_\mathrm{d}$, where $\hat{x} = (1,0)$ is the unit vector in the $x$ direction. This translation of the cluster yields a translation $\mathbf{t} = (r_\mathrm{cm}L/b)\mathcal{R}(\psi+\theta_\mathrm{d}) \hat{x} $ of the moir\'e lattice along the direction $\psi+\theta_\mathrm{d}$ \cite{hermann2012jpcm}, where $\psi$ is the orientation of the moir\'e lattice. Consider also a rotation of the cluster to an orientation $\theta$: this leads not only to the rotation of the moir\'e lattice according to $\tan{\psi}=\sin\theta/(\cos\theta-\rho)$, but also to the shrinkage or expansion of the moir\'e lattice spacing according to $L = b \rho / \sqrt{1+\rho^2-2\rho\cos(\theta)}$, where $\rho=a/b$ is the lattice-spacing ratio of the colloidal cluster and the periodic surface. We index each moir\'e spot by a integer pair $n_1,n_2$. The centers of the moir\'e spots are expressed by $\mathbf{R}_{n_1,n_2}=n_1 \mathbf{A}_1 + n_2 \mathbf{A}_2+\mathbf{t}$, where $\mathbf{A}_1 = L \mathcal{R}(\psi)\hat{x}$ and $\mathbf{A}_2 = L \mathcal{R}(\psi+2\pi/3) \hat{x} $ are the primitive vectors of the moir\'e pattern. We assume that the energy contribution of each moir\'e spot amounts to a Gaussian density profile centered at $\mathbf{R}_{n_1,n_2}$, namely $S_{n_1,n_2}(\theta,\mathbf{r})= \epsilon_\mathrm{M}\exp\left(- \left(\mathbf{r}- \mathbf{R}_{n_1,n_2} \right)^2/(2\lambda^2 L^2)\right)$, where $\epsilon_{\mathrm{M}}=2\epsilon/(\sqrt{3}a^2)$ and $\lambda=0.4875$ determine the strength and width of the Gaussian profile respectively and are here chosen to best replicate the microscopic model and experimental results, see Figs. S4-S6 of the Supplemental Material \cite{supplemental}. By integrating over the area of the cluster oriented at $\theta$ and translated at $\mathbf{r}_\mathrm{cm}$, the interaction energy per particle is   
\begin{equation}
\begin{split}
\langle\epsilon_\mathrm{s}\rangle(\theta,\mathbf{r}_\mathrm{cm}) = &\frac{1}{N}\int \mathrm{d}\mathbf{r}\,h(R_\mathrm{cl}-|\mathbf{r}-\mathbf{r}_\mathrm{cm}|)
 \cdot\Sigma_{n_1,n_2}S_{n_1,n_2}(\theta,\mathbf{r}-\mathbf{t})
\end{split}
\label{eqM7}
\end{equation}
Here the Heaviside function $h()$ cuts the moir\'e spots that are inside the cluster and the summation runs over all integer pairs $n_1,n_2$. The sum of the Gaussian contributions $S_{n_1,n_2}$ can be calculated by means of the Fourier transform
\begin{equation}
\begin{split}
 \sum_{n_1,n_2} S_{n_1,n_2} = \frac{16\pi^3\lambda^2\epsilon_\mathrm{M}}{\sqrt{3}} \int \mathrm{d}\mathbf{q}\,\mathrm{III}(\mathbf{q})
\cdot\exp\left[-\frac{(\lambda L |\mathbf{q}|)^2}{2}\right] \cdot\exp[i \mathbf{q}\cdot (\mathbf{r}-\mathbf{t})], 
\end{split}
\label{eqM8}
\end{equation}
where $\mathrm{III}(\mathbf{q}) = \sum_{\mathbf{Q}} \delta(\mathbf{q} - \mathbf{Q})$  is the Dirac comb, $\mathbf{Q} = m_1 \mathbf{\beta}_1 + m_2 \mathbf{\beta}_2$ a reciprocal moir\'e lattice vector, $m_1,m_2$ integers, $\mathbf{\beta}_1, \mathbf{\beta}_2$ primitive reciprocal moir\'e lattice vectors which satisfy $\mathbf{\beta}_i \cdot \mathbf{A}_j = 2\pi \delta_{ij}$ for $i,j=1,2$. $\delta_{ij}$ is the Kroneker delta function. This way, Eq. (\ref{eqM7}) can be rewritten as:
\begin{equation}
\begin{split}
\langle\epsilon_\mathrm{s}\rangle = -\frac{32\pi^4\lambda^2\epsilon_\mathrm{M}}{\sqrt{3}N} \sum_{\mathbf{Q}} \cos(\mathbf{Q}\cdot\mathbf{t}) e^{-\frac{1}{2} (Q \lambda L)^2} 
\cdot\int_0^\infty \mathrm{d}r\, r\, \mathrm{h}(R_\mathrm{cl}-|\mathbf{r}-\mathbf{r}_\mathrm{cm}|) J_0(Q r). 
\end{split}
\label{eqM9}
\end{equation}
Here $Q=|\mathbf{Q}| = (Kb/L)\sqrt{m_1^2+m_2^2+m_1 m_2}$ with $K=4\pi/(\sqrt{3}b)$, $r = |\mathbf{r}|$, the scalar product $\mathbf{Q}\cdot\mathbf{t}= Kr_\mathrm{cm}[(m_1\cos(\theta_\mathrm{d}-\pi/6) + m_2\sin\theta_\mathrm{d})]$. Since $\textbf{r}_\mathrm{cm}$ is small compared with $\mathbf{t}$, we take the approximation $\mathrm{h}(R_\mathrm{cl}-|\mathbf{r}-\mathbf{r}_\mathrm{cm}|) \approx \mathrm{h}(R_\mathrm{cl}-r)$. The integral in Eq. (\ref{eqM9}) becomes the Hankel transform $\mathcal{F}_\mathrm{Hankel}\,\mathrm{h}(R_\mathrm{cl}-r)=\int_0^\infty \textrm{d}r \,r\, \mathrm{h}(R_\mathrm{cl}-r) J_0(Q r) =J_1(Q R_\mathrm{cl}) R_\mathrm{cl}/Q$. Further substituting $N=2\pi R_\mathrm{cl}^2/(\sqrt{3}a^2)$ in Eq. (\ref{eqM9}), we obtain the energy
\begin{equation}
\begin{split}
 \langle\epsilon_\mathrm{s}\rangle(\theta, \mathbf{r}_\mathrm{cm}) =  -16\pi^3\lambda^2a^2\epsilon_\mathrm{M} \sum_\mathbf{Q} \frac{e^{-\frac{1}{2}\left(Q\lambda L\right)^2}}{QR_\mathrm{cl}} 
 \cdot J_1\left(Q R_\mathrm{cl} \right) \cos(\mathbf{Q}\cdot\mathbf{t}).
\end{split}
\label{eqM10}
\end{equation}
The factor $\exp[-(Q\lambda L)^2/2]/(QR_\mathrm{cl})$ in Eq. (\ref{eqM10}) decays rapidly as $Q$ increases: as an approximation, we consider only the six shortest vectors of $\mathbf{Q}$ in the summation with $(m_1,m_2)=(1,0)$, $(0,1)$, $(-1,1)$, $(-1,0)$, $(0,-1)$, $(1,-1)$,  with length $Q=Ka/L$. This finally leads to Eq. (\ref{eq1}).

For $\delta \neq 0$, the moir\'e spacing $L$ has a finite maximum value $L_\mathrm{max}$. Therefore at cluster size $R_\mathrm{cl} \gg L_\mathrm{max}$, the Bessel function’s oscillation amplitude decays as $(R_\mathrm{cl}/L)^{-0.5}$. By substituting these relations into Eq. (\ref{eq2}) and using Eq. (\ref{eq3}), we see that $\tau_\mathrm{c} \propto R_\mathrm{cl}^{-0.5} \propto N^{-0.25}$. On the other hand, for cluster sizes $R_\mathrm{cl}\ll L_\mathrm{max}$, depinning occurs at the angle $\theta_\mathrm{c}$ when the first ring of moir\'e spots reaches the edge of the cluster, i.e. $L\approx R_\mathrm{cl}$ [see Fig. \ref{fig1}(e,f)]. This leads to $L\approx a/[2\sin(\theta_\mathrm{c}/2)]$, implying $L\sin\theta_\mathrm{c}\approx a$ considering that $\theta_\mathrm{c}$ is generally small (see Fig. S6 of the Supplemental Material \cite{supplemental}). Plugging $L\approx R_\mathrm{cl}$ and $L\sin\theta_\mathrm{c}\approx a$ into Eq. (\ref{eq2}) yields $\tau_\mathrm{c}\propto R_\mathrm{cl}\propto N^{0.5}$. Note that $L_\mathrm{max}\to\infty$ at $\delta=0$, and thus the relation $\tau_\mathrm{c}\propto N^{0.5}$ is valid at any cluster size as shown in Fig. \ref{fig5}(a). Notably, this $\tau_\mathrm{c}\propto N^{0.5}$ scaling matches the $\tau_\mathrm{c} \propto A^{0.5}$ law of macroscopic friction between a rotating disc of area $A$ and a flat surface with uniform friction coefficient \cite{baker}. 

\begin{figure}
  \centering
  \includegraphics[width=1\columnwidth]{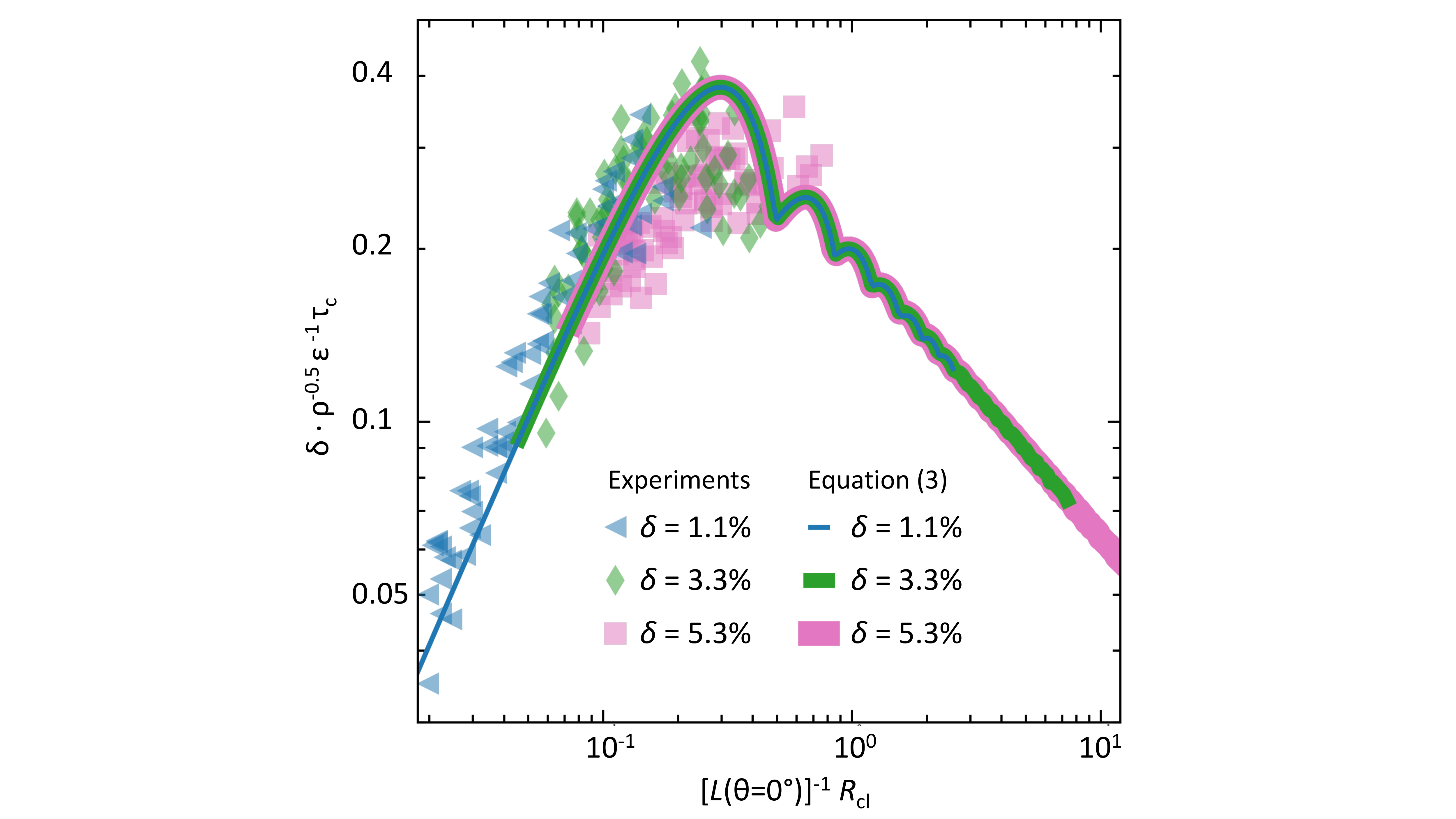}
\caption{Universal size scaling of critical torque. The same curves and experimental data points as in Fig. 2(c) of the main text, but expressed in terms of rescaled cluster size $[ L(\theta=0^{\circ})]^{-1}R_\mathrm{cl}$  and critical torque $\delta \cdot \rho^{-0.5}\epsilon^{-1}\tau_\mathrm{c}$. Observe that the theoretical curves, obtained following Eq. (3) of the main text, overlap perfectly, indicating a universal size-dependence of the critical torque at any mismatched surface, i.e for arbitrary $\delta>0$.}
\label{fig7}
\end{figure}

For the stability diagram in Fig. \ref{fig4}(a) we focus on the $\delta=0$ case at small $\theta$. This leads to $L\approx a/\theta$. For simplicity, we consider only translations along $\hat{x}$, i.e. $\mathbf{r}_\mathrm{cm}=(x,0)$, the potential energy then reads
\begin{equation}
\begin{split}
\langle \epsilon_\mathrm{s}\rangle(\theta,x) =  -\frac{2a\epsilon_\mathrm{eff}}{R_\mathrm{cl}} \theta^{-1} J_1\left(\frac{4\pi R_\mathrm{cl}\theta}{\sqrt{3}a}\right)  \left[ 1+2\cos\left(\frac{2\pi x}{a}\right)\right].
\end{split}
\label{eqM11}
\end{equation}
Fig. S15 of the Supplemental Material \cite{supplemental} reports a comparison between the energy computed in the microscopic model, and the coarse-grained model with Eq. (\ref{eqM11}).

In the presence of an external torque $\tau$ and driving force $\mathbf{F}$ in the $x$ direction, the equilibrium condition satisfies $\partial A / \partial \theta = 0$ and $\partial A / \partial x = 0$, where $A(\theta, x) = \langle\epsilon_\textrm{s}\rangle(\theta, x) - \theta\tau - xF$ is the generalised enthalpy. This leads to  
\begin{equation}
\begin{split}
 \tau = \frac{\partial\langle\epsilon_\mathrm{s}\rangle }{\partial\theta} = \frac{2a\epsilon_\mathrm{eff}}{R_\mathrm{cl}} \theta^{-1} J_2\left(\frac{4\pi R_\mathrm{cl}\theta}{\sqrt{3}a}\right)\left[1+2\cos\left(\frac{2\pi x}{a}\right)\right]. 
\end{split}
\label{eqM12}
\end{equation}

\begin{equation}
\begin{split}
 F = \frac{\partial\langle\epsilon_\mathrm{s}\rangle}{{\partial x}} = \frac{8\pi\epsilon_\mathrm{eff}}{R_\mathrm{cl}} \theta^{-1}J_1\left(\frac{4\pi R_\mathrm{cl}\theta}{\sqrt{3}a}\right)\sin\left(\frac{2\pi x}{a}\right).         
\end{split}
\label{eqM13}
\end{equation}
By differentiating equations (\ref{eqM12}) and (\ref{eqM13}) again with respect to $\theta$ and $x$, we obtain the Hessian matrix $\mathcal{H}=[H_{\theta\theta}, H_{\theta x}; H_{x \theta}, H_{x x}]$:
\begin{equation}
\begin{split}
H_{\theta\theta} = \frac{\partial^2\langle\epsilon_\mathrm{s}\rangle}{\partial\theta^2} = &\frac{2a\epsilon_\mathrm{eff}}{R_\mathrm{cl}} \left[\theta^{-1}J_1\left(\frac{4\pi R_\mathrm{cl}\theta}{\sqrt{3}a}\right) - 3\theta^{-2}J_2\left(\frac{4\pi R_\mathrm{cl}\theta}{\sqrt{3}a}\right)\right] 
\cdot\left[1+2\cos\left(\frac{2\pi x}{a}\right)\right]\\
H_{xx} = \frac{\partial^2\langle\epsilon_\mathrm{s}\rangle}{\partial x^2} = &\frac{16\pi^2\epsilon_\mathrm{eff}}{aR_\mathrm{cl}} \theta^{-1}J_1\left(\frac{4\pi R_\mathrm{cl}\theta}{\sqrt{3}a}\right)\cos\left(\frac{2\pi x}{a}\right)\\
H_{\theta x}= \frac{\partial^2\langle\epsilon_\mathrm{s}\rangle}{\partial x \partial\theta} =&H_{x \theta} =-\frac{8\pi\epsilon_\mathrm{eff}}{R_\mathrm{cl}} \theta^{-1}J_2\left(\frac{4\pi R_\mathrm{cl}\theta}{\sqrt{3}a}\right)\sin\left(\frac{2\pi x}{a}\right) .   
\end{split}
\label{eqM14}
\end{equation}
The sign of the determinant of $\mathcal H$ provides indications about the mechanical stability of the cluster and is shown by the color pattern reported in Fig. \ref{fig4}(a). The stability boundary is defined by $\det\mathcal{H}=0$. This condition provides the critical $(\theta, x)_\mathrm{c}$, reported as a solid line in Fig. \ref{fig4}(a). The corresponding $(\tau, F)_\mathrm{c}$ reported in Fig. \ref{fig4}(b) are obtained by evaluating equations (\ref{eqM12}) and (\ref{eqM13}) at the $(\theta, x)_\mathrm{c}$.

\section{Phenomenological scaling law of static translational friction and our extension to static torsional friction.}
\label{appendscaling}
In a recent work \cite{koren2016prb}, Koren and Duerig (KD) decomposed the static translational friction force $F_\mathrm{static}$ of a crystalline cluster (or flake) interacting with a periodic surface, as follows:
\begin{equation}
F_\mathrm{static} = F_\mathrm{a} + F_\mathrm{e}    
\label{eqM15}
\end{equation}
Here $F_\mathrm{a} = F_\mathrm{a0}R_\mathrm{cl}^{2\beta}$ is the area (or bulk) contribution, $F_\mathrm{e} = F_\mathrm{e0} R_\mathrm{cl}^{\gamma}$ is the edge (or rim) contribution, $R_\mathrm{cl} \propto \sqrt{N}$ is the radius of the cluster, $\beta$ and $\gamma$ are appropriate scaling exponents. According to results of KD, area exponents $\beta=1$ ($\beta=1/4$) are obtained for commensurate (incommensurate) contacts. Edge exponents $\gamma=1$ ($\gamma=1/2$) are obtained for hexagon-shaped (circular-shaped) clusters.
To extend KD’s scaling law, we introduce a position-dependent scaling relation and assume circular-shaped clusters with radius $R_\mathrm{cl}$:
\begin{equation}
\begin{split}
 f_\mathrm{static} = &f_\mathrm{a}(r), ~~ r<R_\mathrm{cl}  \\
 f_\mathrm{static} = &f_\mathrm{e}(R_\mathrm{cl}), ~~ r=R_\mathrm{cl} ,  
 \end{split}
\label{eqM16}
\end{equation}
where $f_\mathrm{a}(r) = f_\mathrm{a0}r^{2\beta-2}$ is the bulk contribution, $f_\mathrm{e}(R_\mathrm{cl}) = f_\mathrm{e0}R_\mathrm{cl}^{\gamma}$ is the edge contribution, $r<R_\mathrm{cl}$ is the distance from the center of the cluster. The integral
 $F_\mathrm{static} = 2\pi \int_0^{R_\mathrm{cl}}r\mathrm{d}r f_\mathrm{a}(r) + 2\pi f_\mathrm{e}(R_\mathrm{cl})$  
of the position-dependent force over the area and the edge recovers Eq. (\ref{eqM15}). 

To evaluate the static torque, and thus rotational friction, we construct the following integral, by multiplying the corresponding position-dependent force in the integrand by the appropriate force arm, namely 
 $\Gamma_\mathrm{static} = 2\pi \int_0^{R_{cl}}r\textrm{d}r [rf_\mathrm{a}(r)] + 2\pi [R_\textrm{cl}f_\textrm{e}(R_\textrm{cl})]$ .
Integration yields
$\Gamma_\mathrm{static} = 2\pi f_\mathrm{a0}R_\mathrm{cl}^{2\beta+1}/(2\beta+1) + 2\pi f_\mathrm{e0} R_\mathrm{cl}^{\gamma+1}$,
or, dividing by the number $N \propto R_\mathrm{cl}^2$ of particles in the cluster, 
\begin{equation}
 \tau_\mathrm{c} = \Gamma_\mathrm{static} / N = \tau_\mathrm{a} N^{(2\beta-1)/2} + \tau_\mathrm{e} N^{(\gamma-1)/2},    
\label{eqM17}
\end{equation}
where $\tau_\mathrm{a}$ and $\tau_\mathrm{e}$ are constants. Even though obtained by assuming circular-shaped clusters, Eq. (\ref{eqM17}) agrees very well with the observed scaling relation even for clusters of different shapes, with the same exponents as obtained by KD.
In a lattice-matched contact ($\delta=0$) where $\beta=1$, the area contribution dominates Eq. (\ref{eqM17}), yielding $\tau_\mathrm{c} \propto N^{0.5}$ regardless of the shape of the cluster. This recovers the small-$N$ scaling we observe in Fig. \ref{fig2}(c).
For a mismatched contact ($\delta>0$) with $\beta=1/4$, the overall scaling of $\tau_\mathrm{c}$ as described by Eq. (\ref{eqM17}) depends on the cluster shape. For hexagon-shaped clusters, the $\gamma=1$ edge contribution becomes the leading term in Eq. (\ref{eqM17}), yielding $\tau_\mathrm{c} \propto N^{0}$, in agreement with the numerical results of Fig. \ref{fig5}(b) and the $\delta=7.3\%$ experimental points of Fig. \ref{fig6}(d). Similarly, for circular-shaped clusters, KD found $\gamma=1/2$ for translational friction. Remarkably, for rotational friction this value leads to the same scaling of the area and the edge contributions, namely $\tau_\mathrm{c} \propto N^{-0.25}$. This scaling is consistent with the large-$N$ envelopes of the results of our numerical simulations and predictions of the coarse-grained analytic formula reported in Fig. \ref{fig5}(a).

\end{document}